%% file: main.tex
\documentclass[11pt]{article} 
\usepackage{smile}
\usepackage{epstopdf}
\usepackage{wrapfig}
\usepackage[colorlinks, linkcolor=blue, anchorcolor=blue, citecolor=blue]{hyperref}
\usepackage[margin=1in]{geometry}
\usepackage[normalem]{ulem}
\usepackage[export]{adjustbox} 
\usepackage{mathtools, cuted}
\usepackage{natbib}
\usepackage{enumerate}
\usepackage{enumitem}

\usepackage{kpfonts}
\DeclareMathAlphabet{\mathsf}{OT1}{cmss}{m}{n}

\SetMathAlphabet{\mathsf}{bold}{OT1}{cmss}{bx}{n}

\newcommand{\commentout}[1]{}

\newtheorem*{theorem*}{Theorem}

\title{\bf Context-Aware Query Rewriting for Improving Users' Search Experience on E-commerce Websites}

\author{
Simiao Zuo$^\dagger$\thanks{Work was done during an internship at Amazon. Correspondence to \texttt{simiaozuo@gatech.edu}.},
Qingyu Yin$^\diamond$, Haoming Jiang$^\diamond$, Shaohui Xi$^\diamond$, \\
Bing Yin$^\diamond$, Chao Zhang$^\dagger$, Tuo Zhao$^\dagger$ \\
$^\dagger$Georgia Institute of Technology \ \ 
$^\diamond$Amazon
}
\date{}

\begin{document}

\maketitle

\begin{abstract}
E-commerce queries are often short and ambiguous. Consequently, query understanding often uses query rewriting to disambiguate user-input queries. While using e-commerce search tools, users tend to enter multiple searches, which we call context, before purchasing. These history searches contain contextual insights about users' true shopping intents. Therefore, modeling such contextual information is critical to a better query rewriting model. However, existing query rewriting models ignore users' history behaviors and consider only the instant search query, which is often a short string offering limited information about the true shopping intent.

We propose an end-to-end context-aware query rewriting model to bridge this gap, which takes the search context into account. Specifically, our model builds a session graph using the history search queries and their contained words. We then employ a graph attention mechanism that models cross-query relations and computes contextual information of the session. The model subsequently calculates session representations by combining the contextual information with the instant search query using an aggregation network. The session representations are then decoded to generate rewritten queries. Empirically, we demonstrate the superiority of our method to state-of-the-art approaches under various metrics. On in-house data from an online shopping platform, by introducing contextual information, our model achieves 11.6\% improvement under the MRR (Mean Reciprocal Rank) metric and 20.1\% improvement under the HIT@16 metric (a hit rate metric), in comparison with the best baseline method (Transformer-based model).
\end{abstract}

\input{0-introduction}
\input{0-background}
\input{0-data-collection}
\input{0-model}

\input{0-experiment}

\input{0-conclusion}

\clearpage
\bibliography{main}
\bibliographystyle{ims}

\end{document}

%% file: 0-introduction.tex
\section{Introduction}

Query rewriting is a task where a user inputs a potentially problematic query (e.g., typos or insufficient information), and we rewrite it to a new one that better matches the user's real shopping intent.
This task plays an important role in e-commerce query understanding, where without proper rewriting, search engines often return undesired items, rendering the search experience unsatisfactory.

\begin{figure}[t!]
    \centering
    \includegraphics[width=0.5\linewidth]{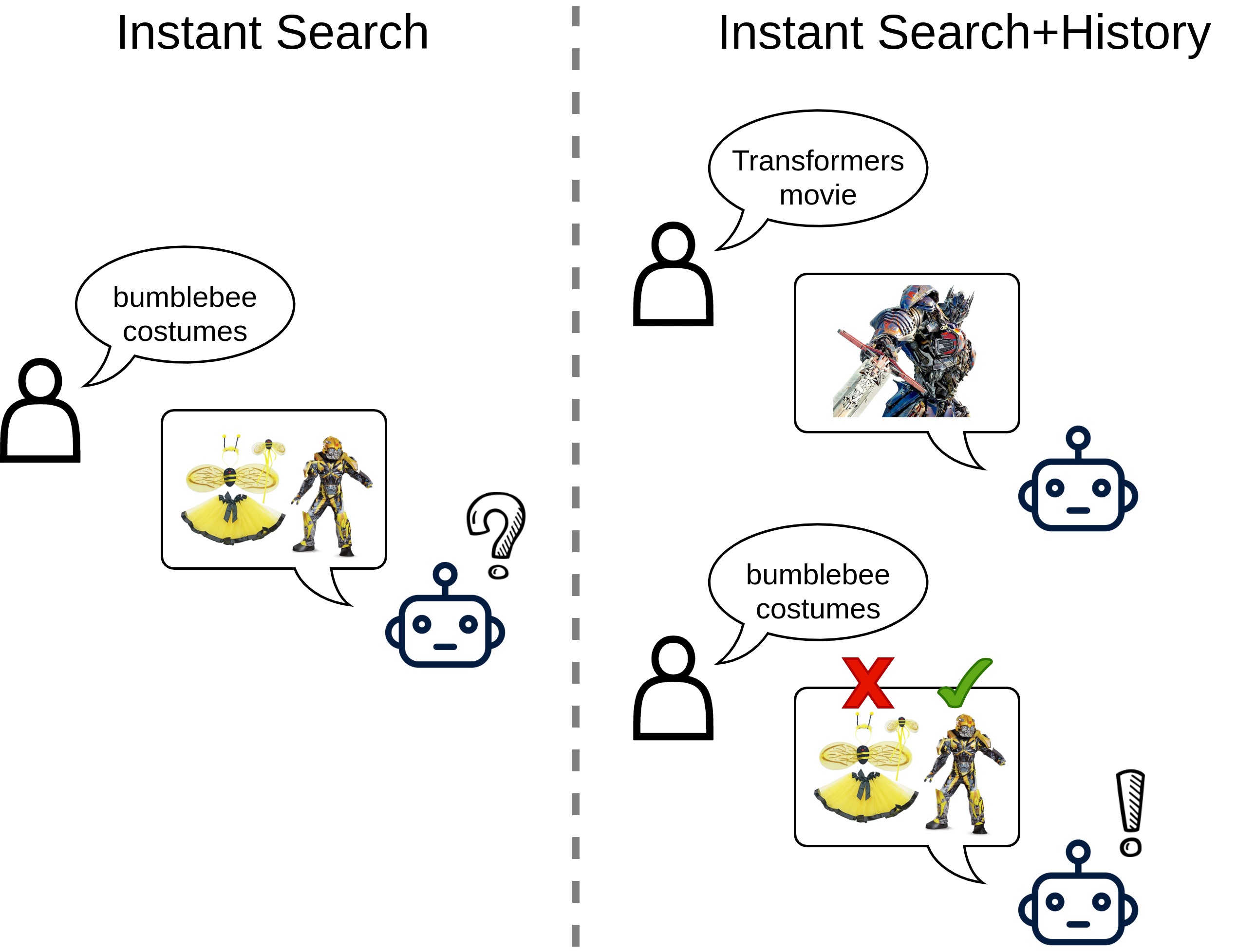}
    \caption{Searching for ``bumblebee costumes'' with (right) and without (left) history searches.}
    \label{fig:bumblebee}
\end{figure}

One major issue that impedes query rewriting is the ambiguity of queries. For example, Figure~\ref{fig:bumblebee} (left) demonstrates searching for ``bumblebee costumes'' without considering search context. From the query alone, it is implausible to tell if the user's intent is for costumes of actual bumblebee (i.e., the animal) or the character from the movie franchise.
This type of ambiguity is common in e-commerce search, where queries are usually short (only 2-3 terms) and insufficiently informative~\citep{he2016learning}. Therefore, it is not possible to disambiguate queries using only the instant search.
A common solution is to use statistical rules to differentiate the possible choices. Specifically, in our example, suppose a total of 100 users entered the ``bumblebee costumes'' query, and 70 of them eventually purchased the movie character costume. When a new user searches for the same query, the recommended products will consist of 70\% movie character costumes and 30\% animal costumes.
This procedure is problematic because each user has a specific intent, i.e., either the movie character costume or the animal costume, but rarely both, which the aforementioned method fails to address.

We propose to explore contextual information from users' history searches to resolve the query ambiguity issue.
Taking the ``bumblebee costumes'' example again, in Figure~\ref{fig:bumblebee} (right), suppose a rewriting model recognizes that the user searched for ``Transformers movie'' earlier, then it could infer that the user's purchase intent is the movie character costume, and hence can remove the input ambiguity.
There have been existing works that utilize search logs for query rewriting. For example, \citet{wang2007learn, wang2008mining} use traditional TF-IDF-based similarity metrics to capture relational information among the user's history searches. These approaches are too restrictive to handle the increasingly complex corpus nowadays. As such, the rewritten queries significantly differ from the original one in intent.
More recently, neural network-based query rewriting algorithms~\citep{he2016learning, xiao2019weakly, yang2019active} are proposed. Most of such approaches employ a multi-stage training approach. Consequently, they involve complicated hand-crafted features or require excessive human annotations for the intermediate features (sometimes both).

To overcome the drawbacks of existing methods, we propose an end-to-end context-aware query rewriting algorithm.
Our model's backbone is the Transformer~\citep{vaswani2017attention}. It is a sequence-to-sequence encoder-decoder model that exploits recent advances of the self-attention mechanism~\citep{bahdanau2014neural}.
In our context-aware model, the Transformer encoder learns representations for individual history queries. The representations are further transformed to carry cross-query relational information using a graph attention mechanism (GAT, \citealt{velivckovic2017graph}).
The GAT computes contextual information of a session based on a session graph, where its nodes contain the history queries and the tokens contained in the history queries.
After obtaining the contextual information from the GAT, it is aggregated with the instant search using an aggregation network. The augmented information is subsequently fed into the Transformer decoder to generate rewritten queries.
Previous works~\citep{tu2019multi, wang2020heterogeneous} that share the same spirit have shown to be effective in various natural language processing tasks.

We highlight that our proposed session graph formulation and the GAT mechanism explicitly models cross-query relations, which is different from existing works.
Previous approaches (e.g., \citep{dehghani2017learning}) capture such relations recursively, which is sub-optimal because such a structure suffers from the ``forgetting'' issue~\citep{hochreiter1997long}, i.e., relation between queries far away will be lost. In contrast, GAT associates any two queries by their contained words, enabling relation-modeling regardless of distance.

Our proposed method improves upon existing works from three aspects.
First, our model does not involve recursion, unlike conventional recurrent neural network-based approaches~\citep{he2016learning, yang2019active, xiao2019weakly}. Our proposed attention-based method can be trained in full parallel and avoids gradient explosion and gradient vanishing problems~\citep{pascanu2013difficulty}, from which existing models suffer. These advantages facilitate training deep models containing dozens of layers capable of capturing high-order information.
Second, our end-to-end sequence-to-sequence learning formulation eliminates the necessity of excessive labeled data. Previous approaches~\citep{yang2019active, xiao2019weakly} require the judgment of ``semantic similarity'', and thus crave for human annotations, which are expensive to obtain. In contrast, our method uses search logs as supervision, which does not involve human effort, and are cheap to acquire.
Third, our method can leverage powerful pre-trained language models, such as BART~\citep{lewis2019bart}. Such models contain rich semantic information and are successful in numerous natural language processing tasks~\citep{devlin2018bert, liu2019roberta, radford2019language}.

We demonstrate the effectiveness of our method on in-house data from an online shopping platform. Our context-aware query rewriting model outperforms various baselines by large margins. Notably, comparing with the best baseline method (Transformer-based model), our model achieves 11.6\% improvement under the MRR (Mean Reciprocal Rank) metric and 20.1\% improvement under the HIT@16 metric (a hit rate metric). We further verify the effectiveness of our approach by conducting online A/B tests.

The remainder of this paper is organized as follows. In Section~\ref{sec:related-work} we review some related works. Section~\ref{sec:setup} describes the problem setup and the data collection process. In Section~\ref{sec:method} we introduce our end-to-end context-aware query rewriting method. Experiments are presented in Section~\ref{sec:experiments}. We conclude this paper in Section~\ref{sec:conclusion}.

%% file: 0-background.tex
\vspace{-0.05in}
\section{Related Works}
\vspace{-0.05in}
\label{sec:related-work}

\noindent $\diamond$
\textbf{Context-based query rewriting }
One line of work uses statistical methods. For example, \citet{cui2002probabilistic, cui2003query} extract probabilistic correlations between the search queries and the product descriptions. Other works extract features that are related to the user's current search~\citep{huang2003relevant, huang2009analyzing}, or from relational information among the user's history searches~\citep{billerbeck2003query, baeza2007extracting, wang2007learn, cao2008context, wang2008mining}.
There are also statistical machine translation-based models~\citep{riezler2007statistical, riezler2010query} that employ sequence-to-sequence approaches.
The aforementioned statistical methods suffer from unreliable extracted features, such that the rewritten queries differ from the original one in intent.

Another line of work focuses on neural query rewriting models~\citep{he2016learning, xiao2019weakly, yang2019active}.
These models adopt recurrent neural networks (RNNs, \citealt{hochreiter1997long, sutskever2014sequence}) to learn a vectorized representation for the user's search query, after which KNN-based methods are used to find queries that yield similar representations. One major limitation is that the rewritten queries are limited to the previously presented ones. Also, these methods often involve complicated and ungrounded feature function designs, e.g., \citet{he2016learning} and \citet{xiao2019weakly} hand-crafted 18 feature functions, or require excessive labeled data~\citep{yang2019active}.
There are other works~\citep{sordoni2015hierarchical, dehghani2017learning, jiang2018rin} that use RNNs for generative query suggestion, but they inherit the weaknesses of RNNs and yield unsatisfactory performance in practice.

Note that \citet{grbovic2015context} construct context-aware query embeddings using word2vec \citep{mikolov2013efficient}. In their approach, an embedding is learned for each distinct query in the dataset. As such, the quality of the learned embeddings rely heavily on the number of occurrences of each query. This method is not applicable to our case because in our dataset, almost all the queries are distinct.

\vspace{0.1in}
\noindent $\diamond$
\textbf{Pre-trained language models }
These models gain increasing attention in natural language processing (NLP). Models such as BERT~\citep{devlin2018bert}, RoBERTa~\citep{liu2019roberta}, and GPT-2~\citep{radford2019language} achieve state-of-the-art performance in various NLP tasks, such as natural language understanding~\citep{he2020deberta} and text classification~\citep{yu2020fine}.
Pre-trained language models are essentially massive Transformer-based neural networks that are trained using enormous open-domain data in a completely unsupervised manner.
When applying these models to downstream tasks, we only need to slightly modify the models instead of training from scratch.
Many popular of these models have either the Transformer encoder (e.g., BERT) or the Transformer decoder (e.g., GPT-2), but not both.
Since we formulate query rewriting as a sequence-to-sequence (seq2seq) learning problem, pre-trained seq2seq models, such as BART~\citep{lewis2019bart} and UniLM~\citep{dong2019unified} are more suitable for the task.

%% file: 0-data-collection.tex
\section{Problem Setup}
\label{sec:setup}

The session data are collected from search logs.
First, we collect all the searches from a specific user within a time window, and we call the searches a ``session''.
After the user purchases a product, the session ends, i.e., we do not consider subsequent queries and behaviors after a purchase happens. This is because after a purchase, the user's intent often change.
Note that different sessions may be collected from different users.

Each session contains multiple searches from the same user. We call the last query in the session the ``target'' query, the second to the last query the ``source'' (or the ``instance) query, and the others the ``history'' queries. The intuition behind this is that because sessions always end with a purchase, the last search (i.e., the target) reflects the user's real intent. When the user enters the second to the last search (i.e., the source), if we can rewrite it to the target query, the user's intent will be fulfilled.

Below is an example of a search session. From the history queries, the user is interested in car related banners/posters.
The source query contains a typo (i.e., ``doger'' is a baseball team) and we should rewrite it to the target query (i.e., ``dodge posters'').
\begin{align*}
    &\text{\textbf{History}: \{dodge banners; mopar banner; mopar poster\}} \\
    &\text{\textbf{Source (Instance)}: dodger posters} \\
    &\text{\textbf{Target}: dodge posters}
\end{align*}


We collect about 3 million (M) sessions, where each session consists of at least 3 history queries, a source query (i.e., the one we need to rewrite), and a target query (i.e., the ground-truth query that is associated with the purchase). We have roughly 18.7M queries, and on average, each session contains 4 history queries.
Query rewriting is consequently formulated as a sequence-to-sequence learning problem. We highlight that per our formulation, we do not need human annotations, unlike existing approaches.


%% file: 0-model.tex
\section{Method}
\label{sec:method}

\begin{figure}[t!]
    \centering
    \includegraphics[width=0.55\linewidth]{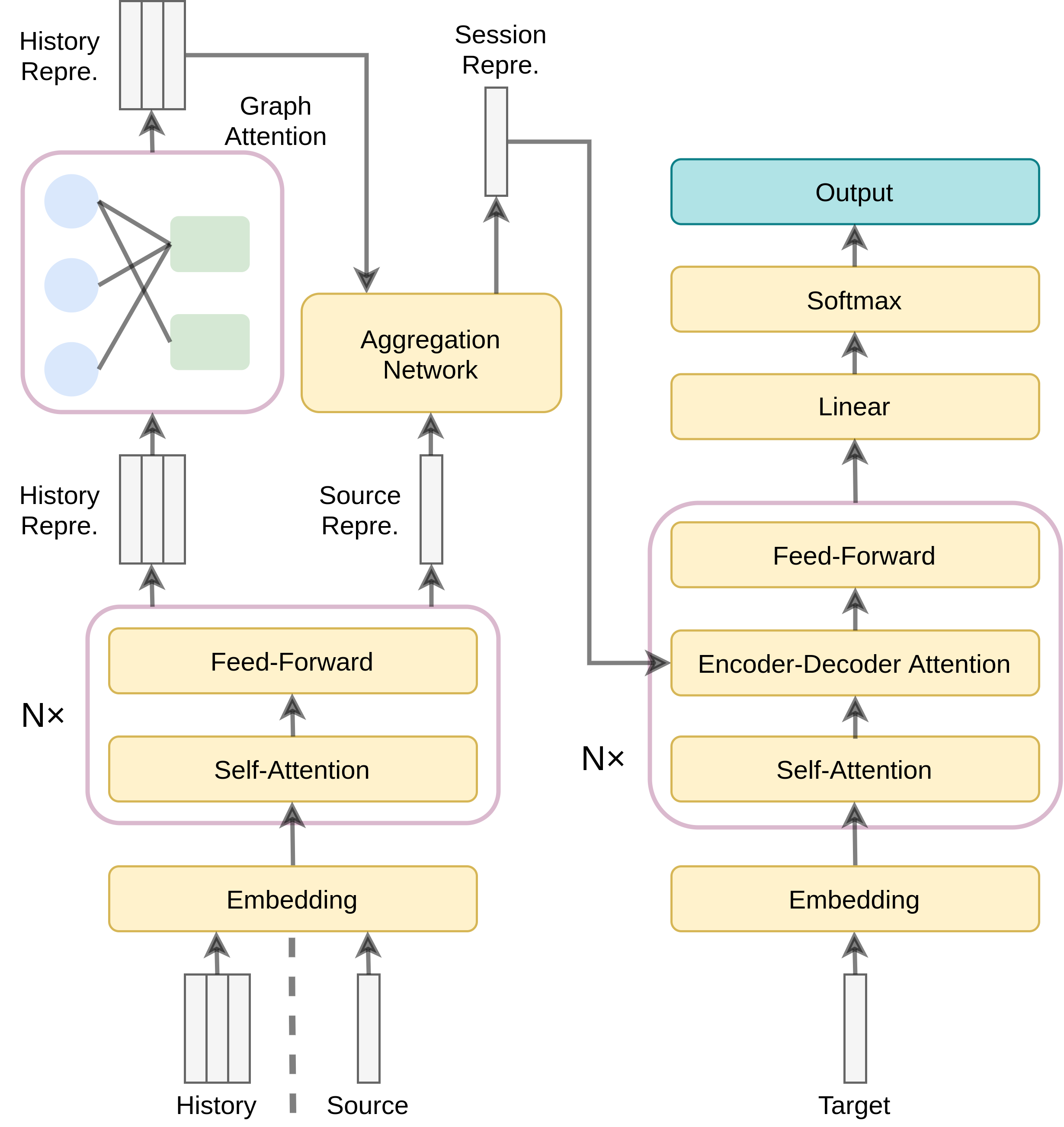}
    \caption{Overview of model.}
    \label{fig:model}
    \vspace{-0.1in}
\end{figure}

Figure~\ref{fig:model} illustrates our context-aware query rewriting model. The model contains four parts: a conventional Transformer~\citep{vaswani2017attention} encoder, a graph attention mechanism~\citep{velivckovic2017graph} that captures the user’s purchase intent, an aggregation network that encodes the history searches, and a conventional Transformer decoder that generates the rewritten query candidates.

\subsection{Transformer Encoder}

For a given source query, we first pad it with a $<$boq$>$ (begin-of-query) token. Then, we pass the padded query through a token embedding layer and a position embedding layer, and we obtain $Y_s \in \RR^{L_s \times d}$. Here $L_s$ is the length of the padded source query, and $d$ is the embedding dimension. Note that the position embedding can either be a sinusoidal function or a learned matrix.

After the initial embedding layers, we pass $Y_s$ through the self-attention module. Specifically, we compute attention output $S$ by
\begin{equation} \label{eq:self-attn}
\begin{split}
    & S = \mathrm{Softmax}\left( \frac{QK^T}{\sqrt{d_k}} V \right), \quad
    \text{where } Q = Y_s W_q, ~ K = Y_s W_k, ~ V = Y_s W_v.
\end{split}
\end{equation}
Here $W_q, W_k \in \RR^{d \times d_K}$, $W_v \in \RR^{d \times d_V}$ are learnable weights. In practice we use multi-head self-attention to increase model flexibility. To facilitate this, different attention outputs $S_1, \cdots, S_H$ are computed using different sets of weights $\{W_q^h, W_k^h, W_v^h\}_{h=1}^H$. The final attention output is then
\begin{align} \label{eq:mult-self-attn}
    S = \left[ S_1, S_2, \cdots, S_H \right] W_o,
\end{align}
where $W_o \in \RR^{H d_V \times d}$ is a learnable aggregation matrix. The attention output is then fed through a position-wise feed-forward neural network to generate encoded representation $H_s \in \RR^{L_s \times d}$ for the source query:
\begin{align} \label{eq:ffn}
    H_s = \mathrm{ReLU}\left( S ~W_\textsubscript{FFN}^1 + b^1 \right) W_\textsubscript{FFN}^2 + b^2.
\end{align}
Here $\{W_\textsubscript{FFN}^1, W_\textsubscript{FFN}^2, b^1, b^2\}$ are weights of the neural network.
Equations \eqref{eq:self-attn}, \eqref{eq:mult-self-attn}, and \eqref{eq:ffn} constitute as an encoder block. In practice we stack multiple encoder blocks to build the Transformer encoder, as demonstrated in Figure~\ref{fig:model}.
For more details about the Transformer architecture, we refer to~\citet{vaswani2017attention}.

For the history queries in this session, we also pad them with $<$boq$>$ tokens.
Suppose that we have $N_h$ padded history queries (recall a session contains multiple history queries), and their respective length is denoted by $L^1_h, \cdots, L^{N_h}_h$. We pad the history queries to the same length, and we obtain the history query matrix $X_h \in \RR^{N_h \times L_h}$, where $L_h = \max \{L^1_h, \cdots, L^{N_h}_h\}$.
Then, following the same procedures as encoding the source query, we pass $X_h$ through the embedding layers and the encoder blocks, after which we obtain the history query representations $U_h \in \RR^{N_h \times L_h \times d}$.

\subsection{Contextual Information from Session Graphs}

After we obtain the history query representations $U_h$, the next step is to refine them.
Such refinement is necessary because the Transformer encoder considers the history queries separately, such that their interactions are not taken into account. However, since each search depends on its previous searches in the same session, modeling cross-query relations are imperative for determining the user's purchase intent.
To this end, we use a graph attention mechanism~\citep{velivckovic2017graph, wang2020heterogeneous} to capture contextual information from $U_h$.

\subsubsection{\textbf{Session Graph Construction}}

\begin{figure}[t!]
    \centering
    \includegraphics[width=0.55\linewidth]{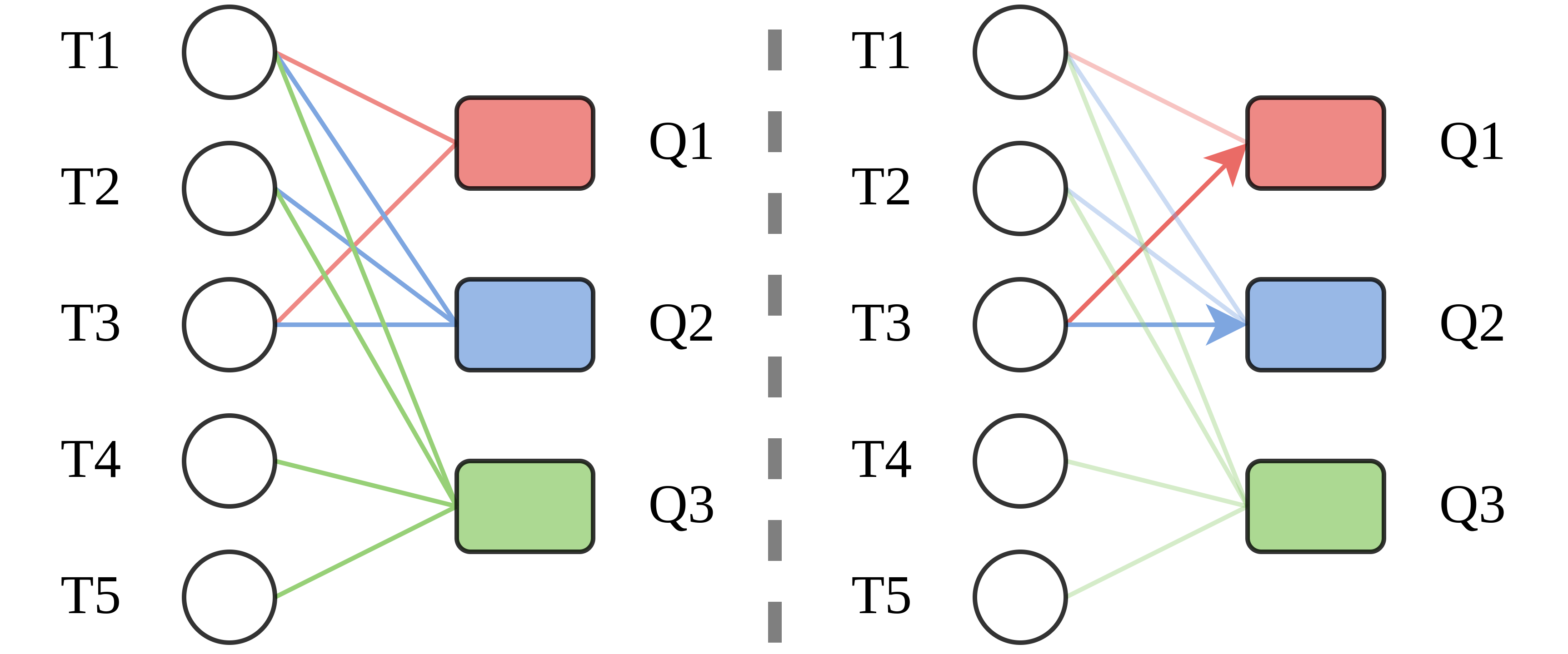}
    \caption{Left: Illustration of a session graph, where ``\textit{T}'' stands for tokens and ``\textit{Q}'' stands for queries. Right: One-step update based on the session graph.}
    \label{fig:gat:sess}
\end{figure}

First we specify how to build a graph for each session, which we call the session graph. 
Suppose we have a session that contains three history queries:
\begin{equation} \label{eq:example}
\begin{split}
    Q_1: \big\{ &\textbf{Search query}: T_1, T_3\}, \\
    Q_2: \big\{ &\textbf{Search query}: T_1, T_2, T_3 \big\}, \\
    Q_3: \big\{ &\textbf{Search query}: T_1, T_2, T_4, T_5 \big\},
\end{split}
\end{equation}
where $Q_1, Q_2, Q_3$ are the three queries, and $T_1, \cdots, T_5$ are the five tokens that appear in the three queries.
Recall Section~\ref{sec:setup} for the problem setup.
Figure~\ref{fig:gat:sess} (left) illustrates the session graph.
In this bipartite graph, the circles are the token nodes ($T_1, \cdots, T_5$); and the rectangles are the query nodes ($Q_1, Q_2, Q_3$).
In our example, the history query representations have size $U_h \in \RR^{3 \times 6 \times d}$, that is, we have 3 queries, and the maximum query lengths is 6 (recall we prepend a $<$boq$>$ token to each query).

\subsubsection{\textbf{Node Representations}}

The next step is to refine the node representations.
Each of the nodes in the session graph has its own representation.

\begin{itemize}
    \item The token representations are simply the corresponding representations of the tokens, extracted from the token embedding matrix.
    \item The query representations are the representations of the $<$boq$>$ token in each padded history query, i.e., the representation of the $Q_1$ query node in Figure~\ref{fig:gat:sess} is found by $U_h[0,0,:] \in \RR^d$. Note that this is akin to BERT, where a $<$cls$>$ token is inserted and its representation is used for classification tasks.
\end{itemize}



Denote $\cG_q = \{q_i\}_{i=1}^{N_q}$ and $\cG_t = \{t_i\}_{i=1}^{N_t}$ the sets of representations for the query and token nodes, respectively.
Here $N_q$ is the number of query nodes and $N_t$ is the number of token nodes. Note that all the node representations have the same size, i.e., $q_i, t_i \in \RR^d$.

\subsubsection{\textbf{Update Node Representations}}

We use a multi-head graph attention mechanism to update the node representations.
For simplicity, denote $N_g = N_q + N_t$ the number of distinct nodes in the session graph, and $\cG = \cG_q \cup \cG_t = \{g_i\}_{i=1}^{N_g}$ the set of all the node representations.

With the above notations, a single-head graph attention mechanism is defined as
\begin{equation} \label{eq:gat}
\begin{split}
    & h_i = g_i + \mathrm{ELU} \left( \sum_{j \in \cN_i} \alpha_{ij} W_v g_j \right), \
    \text{where }
    \alpha_{ij} = \frac{\exp(z_{ij})}{\sum_{\ell \in \cN_i} \exp(z_{i\ell})}, \
    z_{ij} = \mathrm{LeakyReLU} \left(W_a [W_q g_i; W_k g_j] \right).
\end{split}
\end{equation}
Here $\mathrm{ELU}(x) = x \cdot 1\{x>0\} + (\exp(x)-1) \cdot 1\{x\leq 0\}$ is the exponential linear unit, $\cN_i$ denotes the neighbor of the $i$-th node, and $W_a$, $W_q$, $W_k$, $W_v$ are trainable weights.
Note that a residual connection~\citep{he2016deep} is added to the last equation in Eq.~\ref{eq:gat}. This has proven to be an effective technique to prevent gradient vanishing, and hence, to stabilize training.

The session graph only induces attention between nodes that are connected. For example, in Figure~\ref{fig:gat:sess} (right), the model updates $Q_1$ and $Q_2$ using $T_3$, while $Q_3$ is unchanged, i.e., $\cN_{T_3} = \{Q_1, Q_2\}$.

A multi-head graph attention mechanism is then defined as the concatenation of $[h_i^1, h_i^2, \cdots, h_i^K]$, where $K$ is the number of heads, and each of the $h_i$ is calculated via Eq.~\ref{eq:gat}.

The token node representations and the query node representations are updated iteratively. First, we update the token representations ($\cG_t$) using the query representations ($\cG_q$), in order that the tokens acknowledge to which queries they belong. Then, $\cG_q$ is re-computed using the updated version of $\cG_t$, which essentially evaluates cross-query relations, using the token nodes as intermediaries.
Note that the graph attention mechanism (GAT) used in each of the two steps are distinct, i.e., there are two different sets of weights $[W_a, W_q, W_k, W_v]$.

Eventually, we obtain the updated vectorized representations $\{h_i\}_{i=1}^{N_g}$ for all the nodes, and we treat them as the contextual information of the session. 

We remark that the GAT mechanism explicitly models cross-query relations by associating query representations with word representations. Such an approach is fundamentally different from existing methods, where the relations are either ignored (e.g., conventional Transformer attention) or captured via recursion (e.g., RNN-based approaches).

\begin{figure}[t!]
    \centering
    \includegraphics[width=0.5\linewidth]{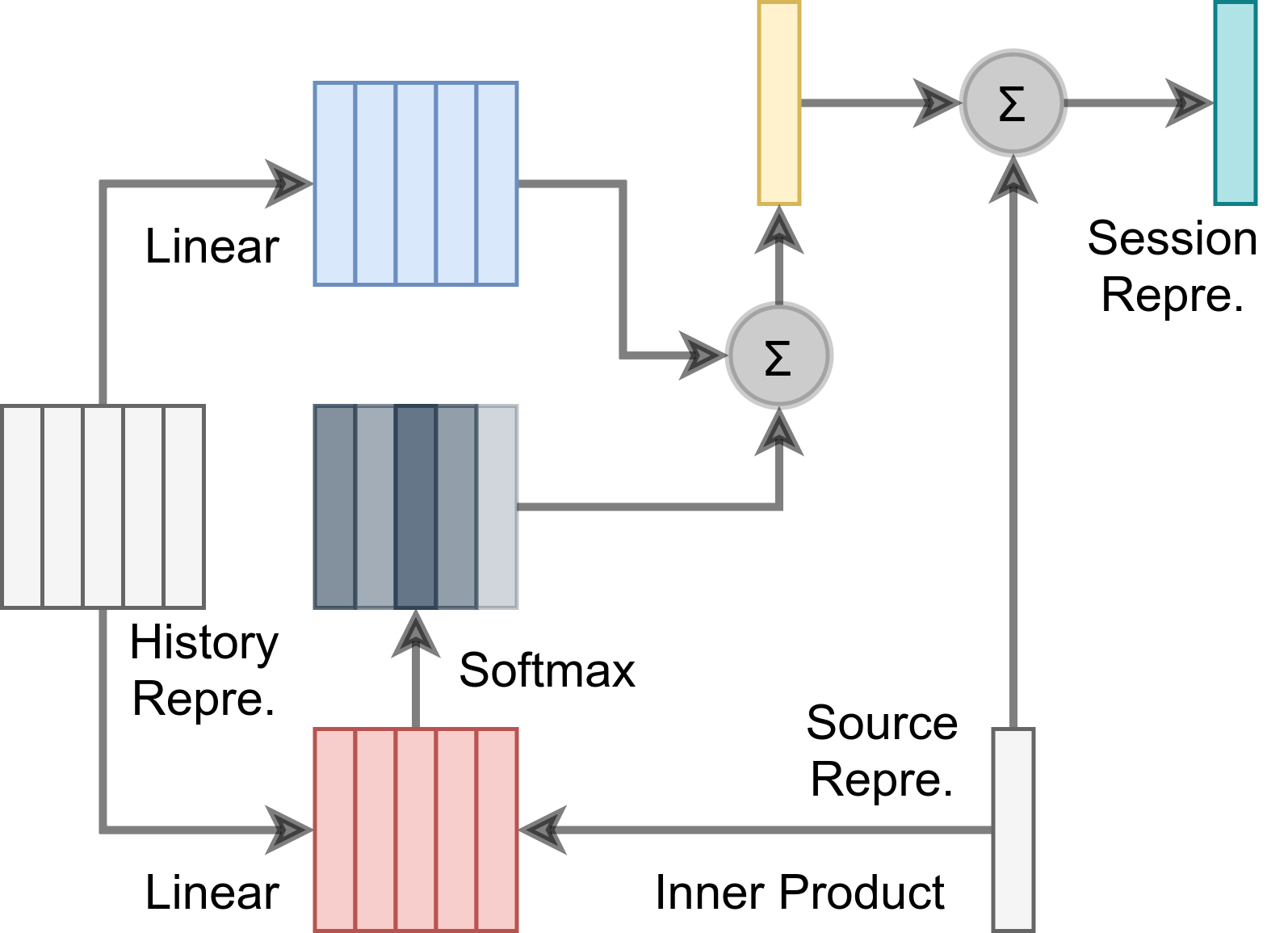}
    \caption{Aggregation network.}
    \label{fig:memory}
\end{figure}

\begin{algorithm}[!t]
\SetAlgoLined
\SetNoFillComment
\caption{Context-aware query rewriting.}
\KwIn{$\cD$: dataset containing sessions;
Initial parameters for the Transformer encoder and the Transformer decoder;
Initial parameters for two graph attention mechanism (Eq.~\ref{eq:gat}): $\mathrm{GAT_{t\rightarrow q}}$, $\mathrm{GAT_{q\rightarrow t}}$;
Initial parameters for the aggregation network (Eq.~\ref{eq:memory});
$K$: the number of updates on the session graph;
$N$: the number of rewritten queries for each session.
}
\KwOut{A list that contains $N$ generated queries for each session in the dataset.}
Rewritten results: $\mathrm{rewritten} = \{\}$\;
\For{each session in $\cD$}{
\tcc{\textbf{Encode input data.}}
Compute source representation $H_s$ and history representation $U_h$ using the Transformer encoder\;
\tcc{\textbf{Apply graph attention.}}
Obtain initial representations $\cG_t^0$, $\cG_q^0$\;
\For{$k = 1\cdots K$}{
    $\cG_t^k = \mathrm{GAT_{q\rightarrow t}}(\cG_q^{k-1}, \cG_t^{k-1})$\;
    $\cG_q^k = \mathrm{GAT_{t\rightarrow q}}(\cG_t^{k}, \cG_q^{k-1/2})$\;
}
Set history representation $\{h_i\}_{i=1}^{N_t+N_h} = \cG_t^K \cup \cG_h^K$\;
\tcc{\textbf{Apply aggregation network.}}
Compute session representation $H_\mathrm{sess}$ from $H_s$ and $\{h_i\}_{i=1}^{N_t+N_h}$ using Eq.~\ref{eq:memory}\;
\tcc{\textbf{Generate rewritten queries.}}
Generate $N$ rewritten queries $\{q_i\}_{i=1}^N$ using the Transformer decoder and a beam search procedure\;
$\mathrm{rewritten} \leftarrow \mathrm{rewritten} \cup \{q_i\}_{i=1}^N$\;
}
\KwOut{The rewritten queries.}
\label{alg:context-rewriting}
\end{algorithm}

\subsection{Session Representation from Aggregation Network}
\label{sec:aggregation}

Recall that we pass the source query through a Transformer encoder and obtain $H_s \in \RR^{L_s \times d}$. The matrix $H_s$ contains representations for all the tokens in the source query. We use that of the prepended $<$boq$>$ token as the representation of the source query, which is denoted $h_s \in \RR^d$.
We adopt an aggregation network to extract useful information with respect to $h_s$ from the contextual information $\{h_i\}_{i=1}^{N_h+N_t}$.
The network employs an attention mechanism that determines to what extent each vector $h_i$ contributes to the source query $h_s$.
Figure~\ref{fig:memory} illustrates the architecture of the aggregation network.
Concretely,
\begin{equation} \label{eq:memory}
\begin{split}
    & H_\textsubscript{sess} = H_s + v, \
    \text{where } v = \sum_{i=1}^{N_g} \alpha_i W_v h_i, \
    \alpha_i = \frac{\exp (z_i)}{\sum_{j=1}^{N_g} \exp(z_j)}, \
    z_i = (W_k h_i)^\top h_s,
\end{split}
\end{equation}
where $W_k$ and $W_v$ are trainable weights. The summation in the last equation in Eq.~\ref{eq:memory} is conducted row-wise, wherein $H_\textsubscript{sess}, H_s \in \RR^{L_s \times d}$, and $v \in \RR^d$.

The matrix $H_\textsubscript{sess}$ serves as the representation of the session. Intuitively, by incorporating the aggregation network, we can filter out redundant information from the session history and only keep the ones pertinent to the source query.

After the Transformer encoder, the graph attention mechanism, and the aggregation network, we obtain $H_\textsubscript{sess}$, the session representation that contains information on both the source query and its history searches. 
Subsequently, $H_\textsubscript{sess}$ is fed into the Transformer decoder to generate rewritten query candidates.

The algorithm is detailed in Algorithm~\ref{alg:context-rewriting}.

%% file: 0-experiment.tex
\section{Experiments}
\label{sec:experiments}

We conduct experiments on some in-house data. Notice that we focus on session-based query reformulation, a scenario that is rare in existing datasets (see Section~\ref{sec:setup} for details).
We implement two methods with different model architectures: \textit{\textbf{Transformer+Aggregation+Graph}} and \textit{\textbf{BART+Aggregation+Graph}}. The first one is constructed in the previous section, and the second one employs a fine-tuning approach instead of training-from-scratch.

\subsection{Training Details}

We use the \textit{Fairseq}~\citep{ott2019fairseq} code-base with \textit{PyTorch}~\citep{paszke2019pytorch} as the back-end to implement all the methods. All the experiments are conducted using 8 NVIDIA V100 (32GB) GPUs.

For training a Transformer model from scratch, we adopt the Transformer-base~\citep{vaswani2017attention} architecture. We use Adam~\citep{kingma2014adam} as the optimizer, and the learning rate is chosen from $\{3\times 10^{-4}, 5 \times 10^{-4}, 1 \times 10^{-3}\}$. We use 4 heads for the multi-head graph attention mechanism, where the head dimension is set to be $128$ (note that the Transformer-base architecture has embedding dimension $512$).

For fine-tuning a BART model, we adopt the BART-base~\citep{lewis2019bart} architecture. We use AdamW~\citep{loshchilov2017decoupled} as the optimizer, and the learning rate is chosen from $\{3\times 10^{-5}, 5 \times 10^{-5}, 1 \times 10^{-4}\}$. Similar to the training from scratch scheme, we adopt 4 heads, each with dimension $192$, for the graph attention mechanism.

For both training-from-scratch and fine-tuning, please refer to\footnote{\url{https://github.com/pytorch/fairseq/blob/master/examples/translation/README.md}} \citet{ott2019fairseq} for more details such as pre-processing steps and other hyper-parameters.

\subsection{Baselines}

The baselines are split into two groups: without pre-training and with pre-training. For the w/o pre-training group, we build the following models:


\vspace{0.05in}
\noindent $\diamond$
\textbf{\textit{Learning to Rewrite Queries (LQRW)}}~\citep{he2016learning} is one of the first methods that applies deep learning techniques to query rewriting. Specifically, the LQRW model combines a sequence-to-sequence LSTM~\citep{hochreiter1997long,sutskever2014sequence} model with statistical machine translation~\citep{riezler2010query} techniques to generate queries. The candidates are subsequently ranked using hand-crafted feature functions. 

\vspace{0.05in}
\noindent $\diamond$
\textbf{\textit{Hierarchical Recurrent Encoder-Decoder (HRED)}}~\citep{sordoni2015hierarchical} employs a hierarchical recurrent neural network for generative query suggestion. The model is a step forward from its predecessors in that HERD is sensitive to the order of queries and the method is able to suggest rare and long-tail queries.

\vspace{0.05in}
\noindent $\diamond$
\textbf{\textit{Transformer}}~\citep{vaswani2017attention} has achieved superior performance in various sequence-to-sequence (seq2seq) learning tasks. To adopt Transformer to query rewriting, we treat the source query as the source-side input, and the target query as
the target-side input. Then we train a model using only these constructed inputs, similar to machine translation. Note that this setting resembles most of the existing works. We adopt the Transformer-base architecture, which contains about 72M parameters.

\vspace{0.05in}
\noindent $\diamond$
\textbf{\textit{MeshTransformer}}~\citep{chen2020incorporating} is a variant of MeshBART, where the pre-trained BART module is replaced by a Transformer and the model is trained from scratch. The method concatenates history queries to the source query in order to integrate contextual information. See the MeshBART method below for details.

\vspace{0.05in}
\noindent $\diamond$
\textbf{\textit{Transformer+Aggregation}} is the model where we use the aggregation network to encode history search queries, i.e., without the graph attention mechanism. Specifically, we first obtain the representations of the source query and the history queries from the Transformer encoder. Then, we extract information related to the source query from the history representations using an aggregation network. Such information is added to the source representation, and we follow a standard decoding procedure using these two factors. See Section~\ref{sec:aggregation} for details.

\vspace{0.1in}
The second group of methods adopt pre-trained language models for query rewriting.

\vspace{0.05in}
\noindent $\diamond$
\textbf{\textit{BART}}~\citep{lewis2019bart} is a pre-trained seq2seq model. We adopt this particular model instead of, for example, BERT~\citep{devlin2018bert} or GPT-2~\citep{radford2019language}, because we treat query rewriting as a seq2seq task. And the aforementioned architectures have either the Transformer encoder (e.g., BERT) or the Transformer decoder (e.g., GPT-2), but not both. In our experiments, BART is fine-tuned in a setting similar to training the Transformer model. We adopt the BART-base architecture in all the experiments, which contains about 140M parameters.

\vspace{0.05in}
\noindent $\diamond$
\textbf{\textit{MeshBART}}~\citep{chen2020incorporating} is a BART-based model that first concatenates the history queries to the source query, and then feeds the concatenated input to a pre-trained BART model for query generation. Note that the original method requires click information. We remove this component as the proposed method do not need such data.

\vspace{0.05in}
\noindent $\diamond$
\textbf{\textit{BART+Aggregation}} is similar to \textit{Transformer+Aggregation}, except we replace the Transformer backbone with the pre-trained seq2seq BART model.

\subsection{Evaluation Metrics}
We use BLEU, MRR (Mean Reciprocal Rank), HIT@1, and HIT@16 to evaluate the query rewriting models. For all metrics except BLEU, we report the gains over the the results calculated by using only source queries.
We remark that MRR, HIT@1, and HIT@16 are more important than BLEU, because MRR and HIT are directly linked to user experience.

We use the BLEU score~\citep{post2018call} as an evaluation metric. This metric is constantly used to evaluate the quality of translation. We adopt it here because similar to machine translation, we formulate query rewriting as a seq2seq learning task. The correlation between the rewritten query and the target query reflects the model's ability to capture the user's purchase intent.

The MRR metric describes the accuracy of the rewritten queries. For each source query in the test set, we generate 10 candidate queries $r_1, \cdots, r_{10}$. Then we search each of these candidates using our production search engine, and we obtain the returned products, of which we only keep the top 32.
Recap that we know the actual product that the customer purchased. 
The next step is to calculate the reciprocal of the actual product's rank for each of $r_1, \cdots, r_{10}$. For example, suppose for $r_1$, the actual purchased product is the second within the 32 returned products, then the score for $r_1$ is $\mathrm{score}_1=1/2=0.5$. The score of the rewritten queries $r_1, \cdots, r_{10}$ is then defined as $\max\{\mathrm{score}_i\}_{i=1}^{10}$. Finally, the score for the query rewriting model is the average over all the source query scores.

We also use HIT@1 and HIT@16 as evaluation metrics. The HIT@16 metric is the percentage that the actual product is ranked within the first 16 products (the first page) when we search the rewritten query. And the HIT@1 metric is similarly defined.

\subsection{Experimental Results}
Table~\ref{tb:exp-transformer} summarizes experimental results.
Recall that in our formulation, we rewrite a source query to a target query. The ``target query'' entry in Table~\ref{tb:exp-transformer} is the performance gain of the ground truth target query, i.e., this entry signifies upper bounds of performance gain that any model can achieve.

We can see that the attention-based models (i.e., BART, MeshBART, Transformer and MeshTransformer) outperforms the recurrent neural network-based approach (i.e., LQRW and HRED).
This is because RNNs suffer from forgetting and training issues. In contrast, Transformer-based models use the attention mechanism instead of recursion to capture dependencies, which has proven to be more effective.
Moreover, by aggregating history searches, \textit{BART+Aggregation} and \textit{Transformer+Aggregation} consistently outperform their vanilla alternatives. Essentially performance of these two methods indicate that integrating history queries into training is critical.
The performance is further enhanced by incorporating the session graphs. Specifically, \textit{Transformer+Aggregation+Graph} achieves the best performance under almost all the metrics.
Notice that the HIT@16 metric gain improves from +15.9 to +20.1 when employing both the aggregation network and the session graph formulation for the Transformer-based models.
We highlight that the graph attention mechanism can directly captures cross-query relations, which is implausible for all the baselines. We can see that this property indeed contributes to model performance, i.e., HIT@16 increases from +17.3 to +20.1 when we equip \textit{Transformer+Aggregation} with the GAT mechanism.

Notice that BLEU is not a definitive metric. For example, the MRR and HIT metrics of HRED are consistently higher than those of LQRW, even though the BLEU score of the former is significantly lower than the latter.
Also, compared with Transformer-based models, the BLEU score is consistently higher when using the BART model as the backbone. This is because a pre-trained language model contains more semantic information. However, the MRR and HIT metrics of the BART-based models are slightly worse than those of the Transformer-based models.

However, the BLEU score is comparable for models with the same backbone. For example, for Transformer vs. \textit{Transformer+Aggregation} vs. \textit{Transformer+Aggregation+Graph}, the BLEU scores are 25.3 vs. 27.2 vs. 28.2. Such a tendency coincides with the online metrics. We observe the same results from BART-based models.


\begin{table*}[t!]
\centering \small
\caption{Experimental results. The results of MRR, HIT@1, and HIT@16 are shown as gain over the source query. The best results are shown in \textbf{bold}.}
\resizebox{1.0\textwidth}{!}{
\begin{tabular}{l|ccc|ccc|c}
\toprule
\textbf{Number of candidates}& \multicolumn{3}{c|}{\textbf{\#Candidates=5}} & \multicolumn{3}{c|}{\textbf{\#Candidates=10}} & \multirow{2}{*}{\textbf{BLEU}} \\ \cline{1-1}
\textbf{Metric} & \ \ \textbf{MRR} \ \ & \ \ \textbf{HIT@1} \ \ & \ \ \textbf{HIT@16} \ \ & \ \ \textbf{MRR} \ \ & \ \ \textbf{HIT@1} \ \ & \ \ \textbf{HIT@16} \ \ &  \\ \midrule
Source Query & 0 & 0 & 0 & 0 & 0 & 0 & --- \\ 
Target Query & +16.1 & +10.6 & +29.0 & +16.1 & +10.6 & +29.0 & --- \\ \midrule
\multicolumn{3}{l}{\textbf{Baseline methods}} \\
LQRW & +3.5 & +2.5 & +6.4 & +6.8 & +4.9 & +12.6 & 29.4 \\
HRED & +4.7 & +3.2 & +8.4 & +8.1 & +5.7 & +14.2 & 25.7 \\
BART & +4.6 & +3.1 & +8.2 & +8.2 & +5.5 & +14.8 & 30.9 \\
Transformer & +4.3 & +2.6 & +9.2 & +8.5 & +5.6 & +15.9 & 25.3 \\
MeshBART & +5.0 & +3.8 & +8.7 & +8.3 & +5.8 & +14.3 & 31.7 \\
MeshTransformer & +4.0 & +2.7 & +8.4 & +8.3 & +5.6 & +15.7 & 25.9 \\  \midrule
\multicolumn{3}{l}{\textbf{Our methods}} \\
BART+Aggregation & +6.3 & +3.9 & +10.9 & +9.7 & +6.4 & +17.1 & 31.9 \\
Transformer+Aggregation & +5.2 & +2.9 & +10.8 & +10.2 & +7.0 & +17.3 & 27.2 \\
BART+Aggregation+Graph & \textbf{+6.9} & \textbf{+4.6} & +11.8 & +10.5 & +7.5 & +17.6 & 32.9 \\
Transformer+Aggregation+Graph \ & +6.6 & \textbf{+4.6} & \textbf{+12.0} & \textbf{+11.6} & \textbf{+8.3} & \textbf{+20.1} & 28.2 \\
\bottomrule
\end{tabular}
}
\label{tb:exp-transformer}
\end{table*}

\input{0-online}

\input{0-analysis}

\input{0-case}

%% file: 0-online.tex
\subsection{Online A/B Test}

To further validate the effectiveness of our approach, we conduct online A/B experiments on a large-scale e-commerce shopping platform with our query rewriting models. For a given search query within a session, we generate one reformulated query using the proposed model, and we feed both the original query and rewritten query into the search system. We run this online query reformulation experiment in the US market.
Experiments are conducted over five days, during which our system processed over $30$ million sessions.
The proposed method improves business metrics in terms of revenue; and also significantly decreases the number of reformulated searches.
This indicates that the rewritten queries better meet customers' shopping intent since customers are able to find their desired products with less number of searches.

%% file: 0-analysis.tex
\subsection{Analysis}

$\diamond$
\textbf{BART vs. Transformer }
Even though BART contains twice the number of parameters compared with Transformer (140M vs. 70M), models fine-tuned on BART yield lower MRR and HIT metrics (Table~\ref{tb:exp-transformer}).
One reason is that publicly available pre-trained models are pre-trained on natural language corpus, but queries are usually short and have distinct structures. This raises doubts on whether current pre-trained models are suitable for the query domain. Indeed, the rich semantic information enables a much better BLEU score (32.9 vs. 28.2), but the MRR and HIT metrics suggest the fine-tuned models' unsatisfactory performance.

Another reason is that in a conventional fine-tuning task, a task-specific head is appended to the pre-trained model, and the head usually contains only a small number of parameters. But in the query rewriting task, both the aggregation network and the graph attention mechanism contain a significant amount of parameters (about 10\% of BART). This is problematic because in fine-tuning, the learning rate is usually small since nearly all the weights are supposed to be meaningful and should not change much. Yet, in our case, we need to properly train a large amount of randomly initialized parameters. Moreover, the aggregation network and the GAT are added inside the pre-trained model (more specifically, they are added to the BART encoder) instead of appended after BART. Essentially this nullifies the pre-trained parameters on the decoder side, imposing additional challenges to the fine-tuning task. Nevertheless, the \textit{BART+Aggregation} model still outperforms the vanilla BART model, and the performance is further improved by adding the GAT (i.e., \textit{BART+Aggregation+Graph}).

\begin{figure*}[t!]
    \centering
    \begin{subfigure}{0.24\textwidth}
         \centering
         \includegraphics[width=\textwidth]{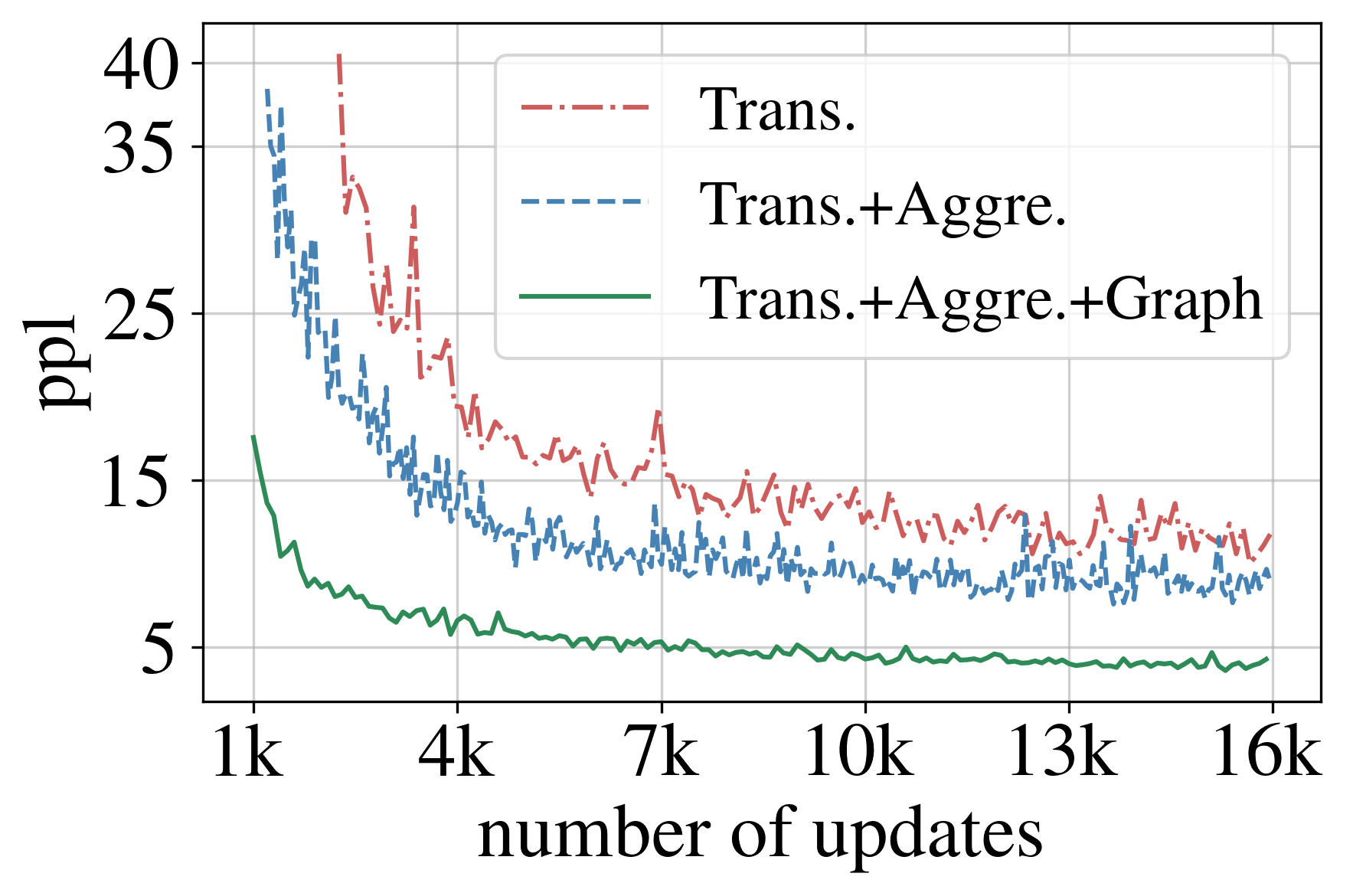}
         \caption{Transformer (train).}
         \label{fig:curve-trans-train}
     \end{subfigure}%
     \begin{subfigure}{0.24\textwidth}
         \centering
         \includegraphics[width=\textwidth]{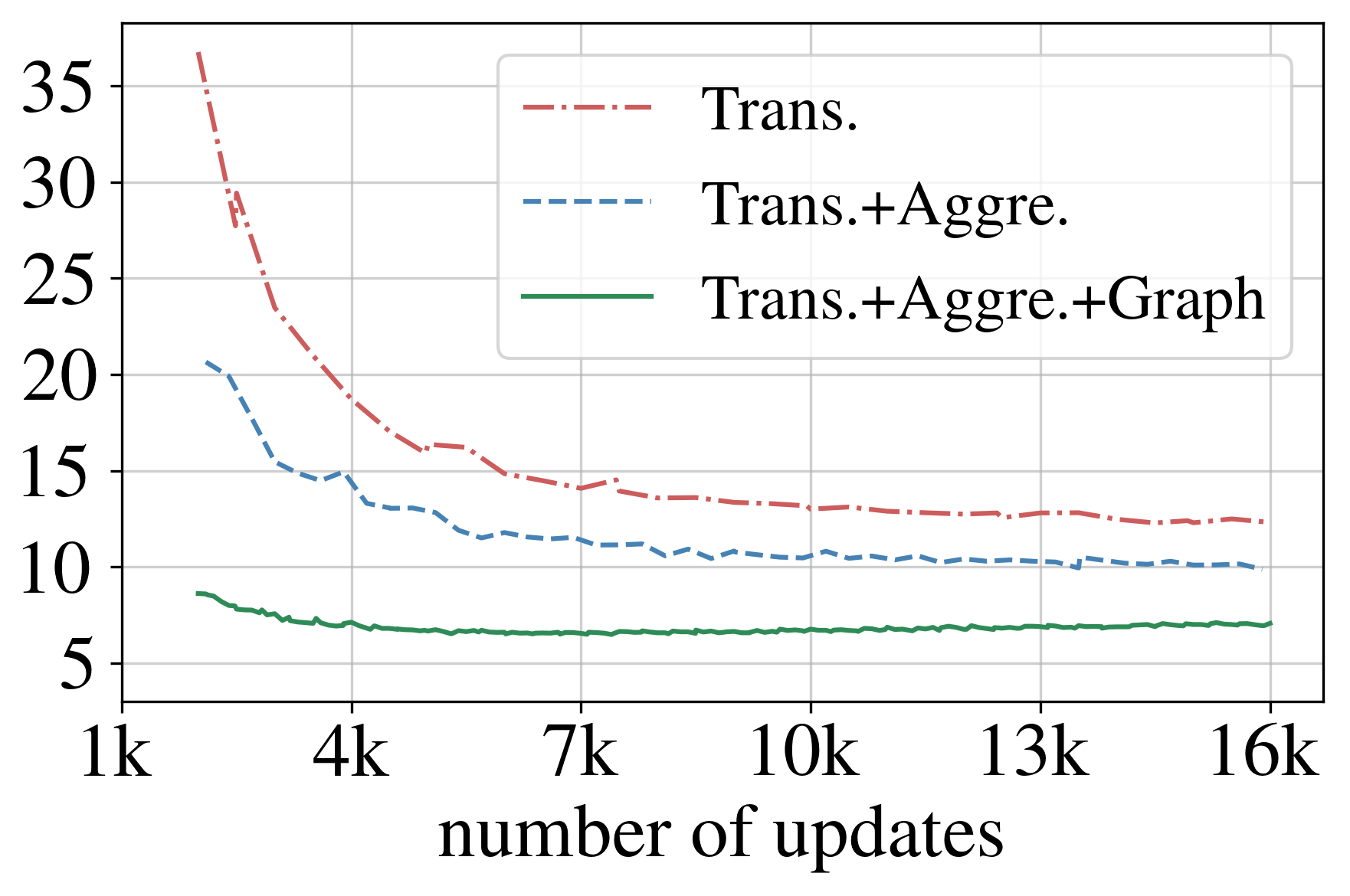}
         \caption{Transformer (valid).}
         \label{fig:curve-trans-valid}
     \end{subfigure}%
     \begin{subfigure}{0.24\textwidth}
         \centering
         \includegraphics[width=\textwidth]{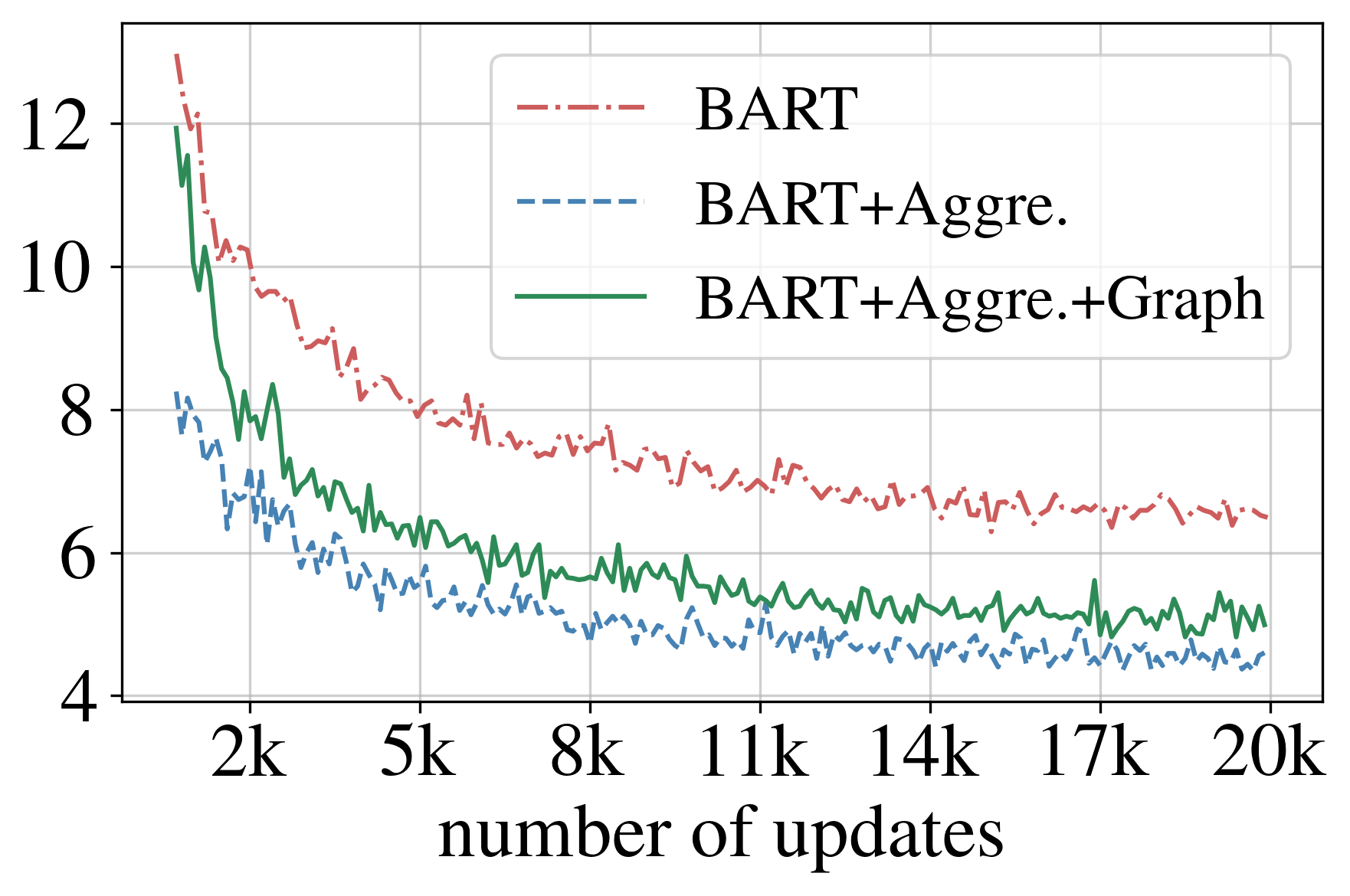}
         \caption{BART (train).}
         \label{fig:curve-bart-train}
     \end{subfigure}%
     \begin{subfigure}{0.24\textwidth}
         \centering
         \includegraphics[width=\textwidth]{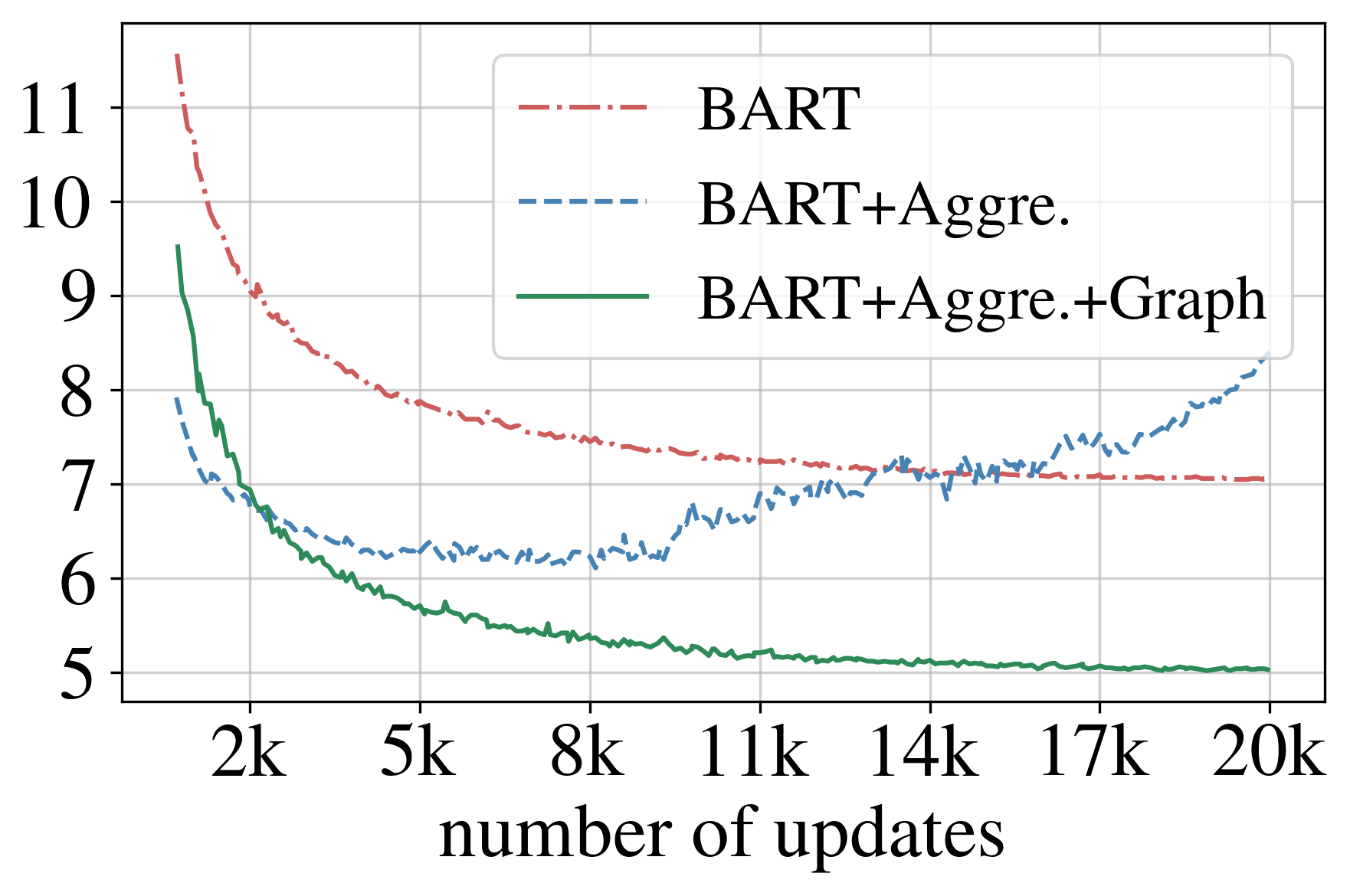}
         \caption{BART (valid).}
         \label{fig:curve-bart-valid}
     \end{subfigure}%
    \caption{Training and validation perplexity using Transformer and BART as backbone.}
    \label{fig:train-curve}
\end{figure*}

\begin{figure*}[t!]
    \centering
    \begin{minipage}[t]{0.6\textwidth}
        \centering
        \includegraphics[width=0.49\textwidth]{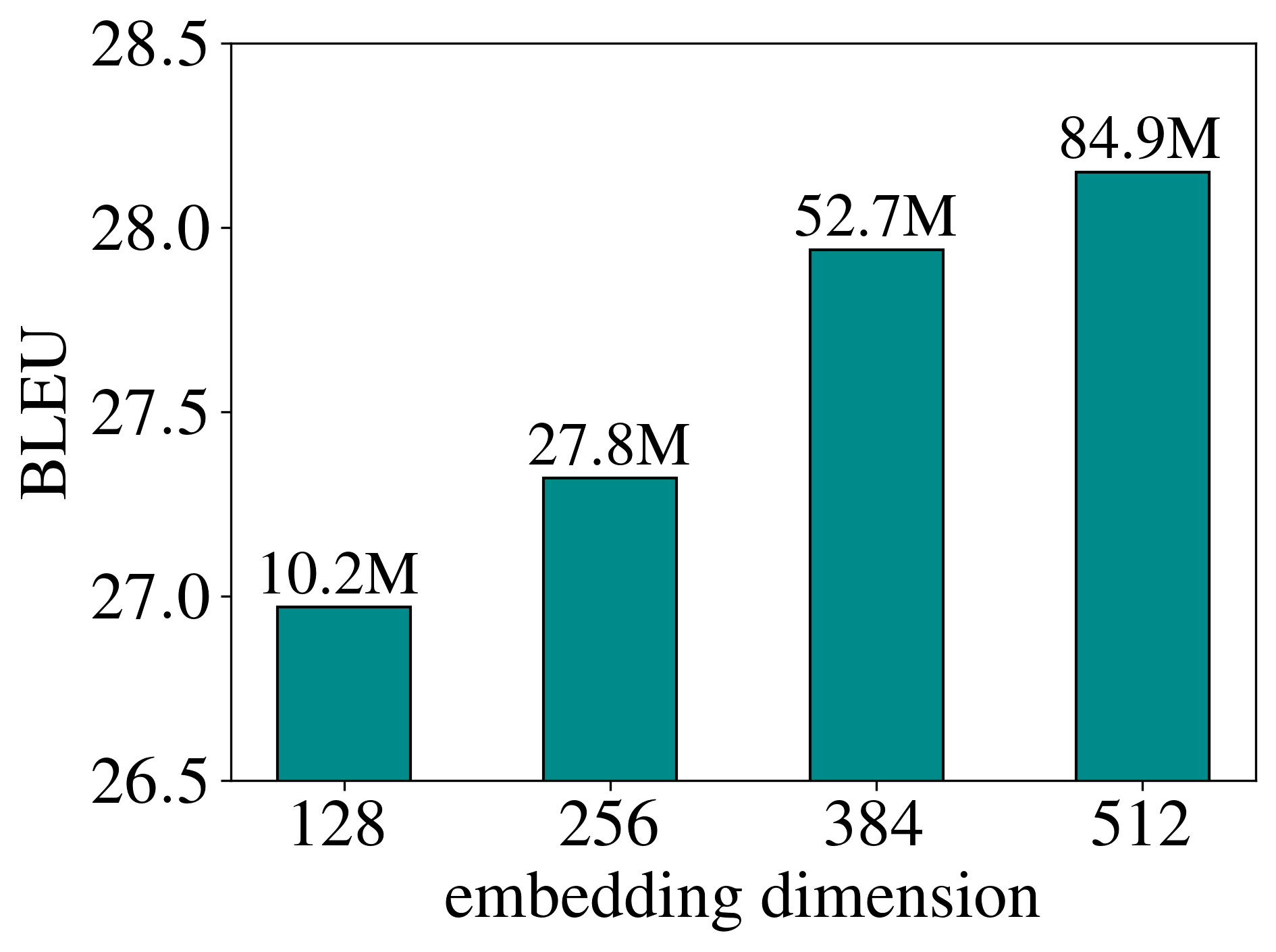}
        \includegraphics[width=0.49\textwidth]{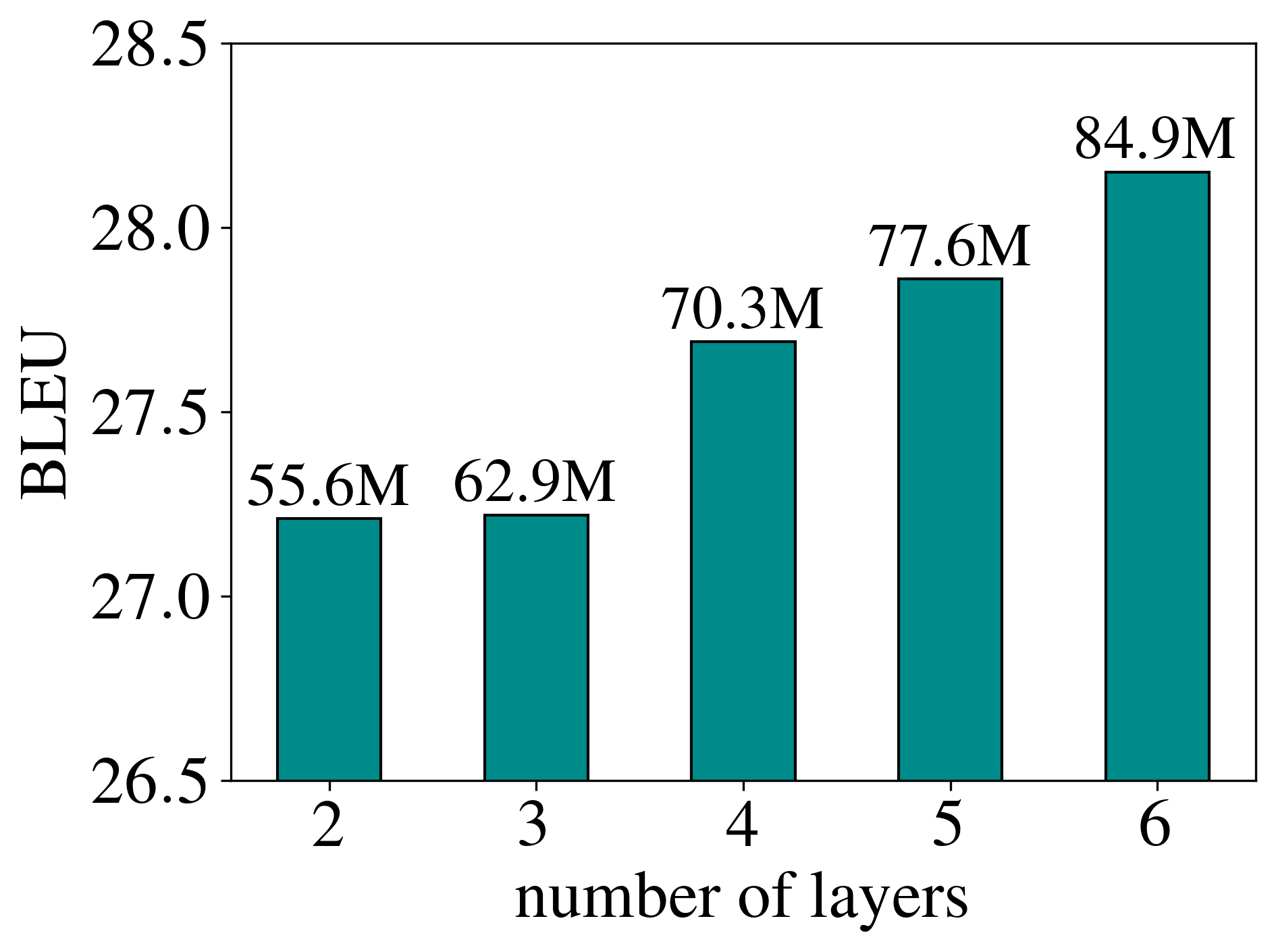}
        \caption{Model performance (in BLEU scores) vs. model size. The model size (in millions of parameters) are shown above the bars.}
        \label{fig:embed}
    \end{minipage}
    \hspace{0.25in}
    \begin{minipage}[t]{0.3\textwidth}
        \centering
        \includegraphics[width=0.96\textwidth]{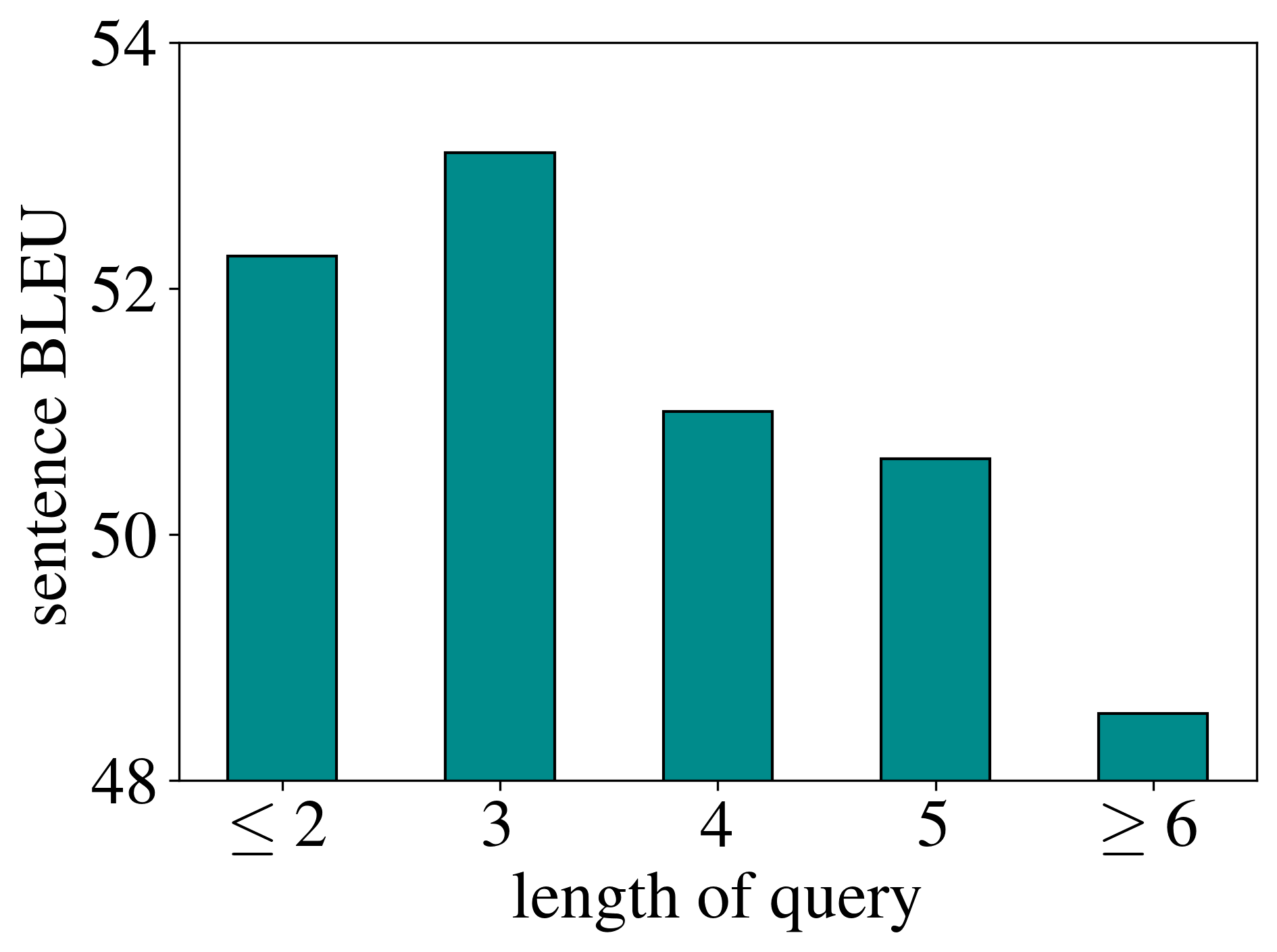}
        \caption{Query length vs. rewriting quality.}
        \label{fig:len-bleu}
    \end{minipage}
\end{figure*}

\vspace{0.1in}
\noindent $\diamond$
\textbf{Training from scratch vs. fine-tuning }
Figure~\ref{fig:train-curve} plots the training and validation perplexity (ppl) of the training-from-scratch approach and the fine-tuning approach. From Figure~\ref{fig:curve-trans-train} and Figure~\ref{fig:curve-trans-valid}, we can see that by employing the aggregation network, \textit{Transformer+Aggregation} fits the data better and exhibits enhanced generalization. The training and validation ppls are further significantly improved by incorporating the graph attention mechanism, i.e., by using \textit{Transformer+Aggregation+Graph}, we achieve even better performance.

Notice that in Figure~\ref{fig:curve-bart-train}, \textit{BART+Aggregation} outperforms \textit{BART+Aggregation+Graph} in terms of training ppl, which is different from the training-from-scratch approach. As indicated by Figure~\ref{fig:curve-bart-valid}, \textit{BART+Aggregation} shows clear sign of over-fitting. This is because even though pre-trained language models contain rich semantic information, much of it is considered ``noisy'' for query rewriting. Thus feature enhancement initiated by the graph attention mechanism is needed.

\vspace{0.1in}
\noindent $\diamond$
\textbf{Model size vs. performance} Figure~\ref{fig:embed} illustrates the relation between model size and performance, where we decrease the embedding dimension (correspondingly the FFNs' hidden dimensions) and the number of layers. We can see that even with $1/8$ of the parameters, model performance does not decrease much. Moreover, our model is more than $20\%$ smaller than a BERT-base model (85M vs. 110M), rendering online deployment more than possible.

\vspace{0.1in}
\noindent $\diamond$
\textbf{Query length vs. performance} Figure~\ref{fig:len-bleu} demonstrates model performance regarding length of the instant query. We can see that the BLEU score gradually decreases when the length increases. This is because long queries are often very specific (e.g., down to specific models or makes), making the rewriting task harder.

%% file: 0-case.tex
\subsection{Case Studies}

\newcolumntype{C}{@{\hskip3pt}c@{\hskip3pt}}
\newcolumntype{L}{@{\hskip3pt}l@{\hskip3pt}}

\begin{table}[tb!]
\centering \small
\caption{Two examples of context-aware query rewriting with and without context.}
\begin{tabular}{l|c}
\toprule
\textbf{Example 1} & dodge led sign; \\
\multirow{3}{*}{History} & dodge banners; \\
& mopar banner; \\
& mopar poster \\ \midrule
Source & dodger posters  \\ \midrule
Target & dodge posters \\ \midrule
Rewritten w/o context & dodger flag \\ \midrule
Rewritten w/ context & dodge poster \\ \bottomrule \bottomrule
\textbf{Example 2} & samsung galaxy case; \\
\multirow{2}{*}{History} & samsung galaxy a11 case; \\
& samsung a11 case \\ \midrule
Source & samsung galaxy a7  \\ \midrule
Target & samsung galaxy a7 case \\ \midrule
Rewritten w/o context & samsung galaxy a7 charger \\ \midrule
Rewritten w/ context & samsung galaxy a7 case \\
\bottomrule
\end{tabular}
\label{tb:rewrite-demo}
\end{table}

\noindent $\diamond$
\textbf{Advantages of leveraging history information }
Two examples are shown in Table~\ref{tb:rewrite-demo}.
The first example is error correction. In the example, the customer wishes to purchase dodge (a car brand) posters, but she mistakenly searches for dodger (a baseball team) posters. Without history information, it is impossible to determine the customer’s true intent. However, by looking at session histories, we find that all the previous searches are related to automobiles (e.g., dodge and mopar), and therefore the query should be rewritten to ``dodge posters''. Our model successfully captures this pattern.
Notice that the rewritten query without leveraging context does not match the user's intent.

The second example is keyword refinement. In the example, by looking at the history searches, it is obvious that the customer wishes to find phone cases, instead of phones. However, this intent is impossible to capture by using only the source query. Our model automatically adds the keyword ``case'' to the source query and matches the target query. On the other hand, without the context information, the rewritten result is not satisfactory.

\vspace{0.1in}
\noindent $\diamond$
\textbf{Diversity of query generation }
Table~\ref{tb:diversity} demonstrates two examples.
In the first example (the left three columns), notice that our model can grep information from history queries, e.g., ``iphone 11 case sailor moon'', and can delete keywords that are deemed insignificant or too restrictive, e.g., ``iphone 11 case leopard'' instead of ``snow leopard''.
Also, our model can effectively capture domain information. For example, some of the history query keywords (e.g., pokemon, eevee) are often described as ``cute'', and our model recommends this keyword. All the history keywords are from Japanese anime series, therefore our model suggests another popular character, ``totoro''. Additionally, the ``disney'' and ``disney princess'' keywords are generated based on the interest to virtual characters. Finally, notice that the likelihood of all the suggested queries is similar, which means our model cannot single out a significantly better query than the others. Therefore our model generated a diverse group of queries.

In the second example (the right two columns), the generated query successfully matches the target query. Note that the top two generated queries have high likelihood, and the likelihood decreases drastically as the suggested queries become more and more implausible. In this example, the first query is 172\% more likely than the tenth query, whereas this number is only 41\% in the previous example. This suggests that our model can differentiate between good quality suggestions and poor quality alternatives.

\begin{table*}[tb!]
\centering \small
\caption{Two examples of generated queries and their associated likelihood.}
\resizebox{1.0\textwidth}{!}{
\begin{tabular}{l|c|c||c|c}
\toprule
Type & Query & Likelihood & Query & Likelihood \\ \midrule \midrule
\multirow{4}{*}{History} & iphone 11 pro case pokemon; & & \multirow{2}{*}{colorado 2005 tail lights;} \\ 
& iphone 11 pro case eevee; & \multirow{2}{*}{---} & \multirow{2}{*}{colorado 2005 door} & \multirow{2}{*}{---} \\
& iphone 11 pro case hetalia; & & \multirow{2}{*}{colorado 2005 accessories} \\ 
& iphone 11 pro case sailor moon & & \\ \midrule
Source & iphone 11 pro case snow leopard & --- & colorado headlights & --- \\ \midrule
Target & iphone 11 pro case tiger & --- & colorado 2005 headlights & --- \\ \midrule \midrule
\multirow{10}{*}{Rewritten}
& iphone 11 pro case disney          & 0.497 & 2005 colorado headlights     & 0.566 \\ 
& iphone 11 pro case sailor moon     & 0.492 & colorado headlights 2005     & 0.458 \\ 
& iphone 11 pro case harry potter    & 0.445 & colorado headlights led      & 0.357 \\ 
& iphone 11 pro case                 & 0.440 & colorado headlights assembly & 0.301 \\ 
& iphone 11 pro case cute            & 0.419 & colorado tail lights         & 0.289 \\ 
& iphone 11 pro case leopard         & 0.391 & colorado headlights housing  & 0.237 \\ 
& iphone 11 pro case clear           & 0.379 & colorado led headlights      & 0.234 \\ 
& iphone 11 pro case disney princess & 0.372 & 2004 colorado headlights     & 0.230 \\ 
& iphone 11 pro case pink            & 0.364 & colorado 2004 headlights     & 0.214 \\ 
& iphone 11 pro case totoro          & 0.353 & colorado headlights 2004     & 0.208 \\
\bottomrule
\end{tabular}
}
\label{tb:diversity}
\end{table*}

%% file: 0-conclusion.tex
\section{Conclusion and Discussions}
\label{sec:conclusion}

We propose an end-to-end context-aware query rewriting model that can efficiently leverage user's history behavior. Our model infers a user's purchase intent by modeling her history searches as a graph, on which a graph attention mechanism is applied to generate informative session representations. The representations are subsequently decoded into rewritten queries.
We conduct experiments using in-house data from an online shopping platform, where our model achieves 11.6\% and 20.1\% improvement under the MRR and HIT@16 metrics, respectively.
Online A/B tests are also conducted to further demonstrate the effectiveness of the proposed context-aware query rewriting algorithm.

Our proposed session graph is flexible, and can be extended to incorporate more information. In this paper, we present a bipartite graph, which contains words and queries. Additional components can be added as extra layers to the session graph. For example, we can add product information such as categories to the session graph, which will turn the current bipartite graphs (word and query) to 3-partite graphs (word, query and product).